\documentstyle[aps,epsfig,preprint,tighten,float]{revtex}
\newcommand{\be}{\begin{equation}}
\newcommand{\ee}{\end{equation}}

\begin{document}
\title{The Dielectric Breakdown Model at Small $\eta$: Pole Dynamics}
\author{M. B. Hastings}
\address{Physics Department, Jadwin Hall\\
Princeton, NJ 08544\\
hastings@feynman.princeton.edu}
\maketitle
\begin{abstract}
We consider the dielectric breakdown model in the limit $\eta\rightarrow 0^+$.
A differential equation describing the surface growth is derived; this
equation is KPZ plus a term causing linear instability, and includes
a short-distance regularization similar to a viscosity.
The equation exhibits an interesting
dynamics in terms of poles, permitting us to derive a large family of
solutions to the equation of motion.  In terms of poles, we find all stationary
configurations of the surface, and analytically calculate their stability.
For each value of the viscosity, only one
stable configuration is found, but this configuration is
nonlinearly unstable in the presence of an exponentially small
amount of noise.  
The present approach may be useful in understanding
the dynamics for finite $\eta$, in particular the DLA model, in terms of 
perturbations to the infinitesimal $\eta$ problem.
\end{abstract}
\section{Introduction}
The problem of Laplacian growth in two dimensions has been much studied
since the introduction of the model of Diffusion Limited Aggregation 
(DLA)\cite{dla}.  The model is known to give rise to complex, branching
structures, with a non-trivial fractal dimension.

In radial geometry, 
the DLA model is defined by starting with a seed particle located
in the two-dimensional plane, and letting
additional particles arrive from infinity by a series of random walks.
Equivalently, one can define a Laplacian field, $\phi$, subject
to $\nabla^2\phi=0$, with boundary 
conditions such that $\phi=0$ on the cluster boundary and
$\phi(r)\propto {\rm log}(r)$ as $r\rightarrow\infty$.  Then, the DLA
model is defined by the statement that the probability of adding a
particle at any point along the cluster boundary is proportional to
the normal derivative of $\phi$, where one solves for $\phi$ after every
growth step.  

One can also define the problem in strip geometry, which
is what we will use throughout this paper.  Here, the growth is
confined to a strip, which is periodic in the horizontal direction,
with the growth happening vertically.
The initial, seed cluster is taken to be a flat surface, and
random walkers are released from infinity, above the strip.
In the strip geometry, one finds fractal clusters when the ratio of
the width of the strip to the walker size diverges.  We will refer to
this as the IR limit of the problem.

A useful generalization of the DLA model is the Dielectric Breakdown
Model (DBM)\cite{dbm}, which includes an additional free parameter $\eta$.  
Then, the probability to add a particle at a given point along the cluster
boundary is taken to be the normal derivative at that point, raised to
the power $\eta$.  For large $\eta$, growth
occurs largely at the tips of the cluster, and the branches of the
cluster become very long and thin, with a fractal dimension approaching
1 as $\eta\rightarrow\infty$.  For small $\eta$, the cluster becomes less
branched, with a dimensionality approaching 2 in the limit $\eta\rightarrow 0$,
where the dynamics reduces to an Eden model\cite{eden}.

The idea behind the present work is to examine the DBM in the limit of
infinitesimal, positive $\eta$.  For any positive $\eta$, in strip
geometry, the flat surface is unstable.  However,
in the limit $\eta\rightarrow 0^{+}$, the fluctuations about the
flat surface become small.  In the next section, we will show how to
rescale the fluctuations in the surface to obtain a non-trivial growth
equation describing the $\eta\rightarrow 0^{+}$ limit.

What we hope to do is to describe the $\eta\rightarrow 0^{+}$ limit
sufficiently well that the problem at finite $\eta$ can be handled
by some kind of perturbation theory; we are still far from this
goal, but this paper represents a first step in that direction.
Note, though, that for any non-infinitesimal
$\eta$, the growth model gives rise to fractal structures with
nontrivial scaling; this implies that the perturbation theory will
give rise to an IR divergence and require RG techniques.  
We will discuss this more in the
conclusion.
The nontrivial scaling at finite $\eta$ should be associated with a crossover 
length, such that on scales
less than this length the surface would be self-affine, like KPZ, but on
scales greater than this length it would be fractal.

We will find that our problem exhibits an interesting pole
dynamics.  We derive a large family of
solutions to the equation of motion using the pole dynamics, and
find all stationary
configurations of the surface, as well the stability of these configurations.
We will find that the $\eta\rightarrow 0^{+}$ limit of the problems
contains many of the essential physical features of the finite $\eta$ problem:
non-linear instability for exponentially small bare noise and competition
between fingers.  
Finally, the dynamics of moving
poles that we will uncover is very similar to the branched growth
dynamics of Halsey and coworkers\cite{halsey}; the 
location of the poles that we find
will correspond to points on the surface of the cluster separating
branches in that approach.  In a future work\cite{me}, we will elaborate more
on the relationship of pole dynamics to branches of the cluster, and
present a more complete
theory of the statistical properties of the surface when the system
is destabilized by noise.

In addition to being a useful limit of the DBM,
the  $\eta\rightarrow 0^{+}$ model should be useful for certain other
physical situations.  The instability term that we will derive is to
some extent reminiscent of a Mullins-Sekerka instability\cite{ms}, found in
solidification problems.  While the DLA model describes growth
processes limited by a diffusive field, the KPZ model describes growth
of a surface normal to itself, with no diffusive field.  One
expects that the equations we derive will be valid in describing growth
of a surface normal to itself, with some weak coupling to a diffusive
field, such as would be found in solidification problems.
\section{Differential Equation for Surface Growth}
In this section, we derive the differential equation describing the surface 
growth in the
$\eta\rightarrow 0$ limit.  First we introduce the continuum equation of
motion for the DBM, and then we take the appropriate
limit.  We rescale the fluctuations in the height of the surface in this
limit.  We regularize the equation on a short distance, and argue that this
regularization corresponds to a finite walker size in the DBM.  Finally
we discuss some basic scaling properties of the resulting equation and discuss
how to include noise in the dynamics.

Although some time will be spent deriving the equation of motion in
the $\eta\rightarrow 0$ limit, the final result is quite simple.
It will turn out to just be KPZ dynamics\cite{KPZ}, describing the
surface at $\eta=0$, with an additional linear
instability due to the non-vanishing $\eta$.

Consider the dielectric breakdown model.  As discussed above,
there are two different kinds of
geometries commonly used.  One possibility is to grow out radially.  Another
possibility is to grow upwards on a strip of finite width, using periodic 
boundary conditions.  We will
use the strip geometry throughout this paper, although in later sections we will
occasionally discuss limiting cases of the strip geometry, in which the
width of the strip becomes large, and the growth becomes the same as growth
upwards from the real line.  For the strip boundary conditions, we will use
a coordinate $x$, periodic with period $2\pi$, to parametrize the cluster.

We will use complex variables to describe the coordinates in the two-dimensional
strip in which growth occurs.  The real part of the complex variable
will represent the horizontal position, while the imaginary part represents
the vertical position.  Growth will occur in the positive imaginary direction.

We will parametrize the boundary of the cluster by using a function $F(x)$, such
that $F(x)$ yields the cluster boundary, for $x=0...2\pi$.  $F(x)$ is taken
to be analytic, and one-to-one for $x$ with positive imaginary part.  In
the limit that $x\rightarrow +i \infty$, we require that $F(x)\rightarrow x$.
This implies that $F(x)$ can be written as
\be
F(x)=x+\sum_{j=0}^{\infty} F(j) e^{i j x}
\ee
Throughout, we will use the symbols $j,k,...$ to refer to modes in
Fourier space, while $x$ will be used to refer to real space.  The
function $F(j)$ vanishes for $j<0$.

The surface does not contain overhangs in our limit,
although overhangs are essential for the fractal clusters at
finite $\eta$.   The absence
of overhangs permits us to describe the surface by its height, a
function $h(x)$.  
If we expand $h(x)$ in powers of $F$ we obtain
\be
\label{height}
h(x)=
{\rm Im}(F(x))+...
\ee
For small $F$, the higher order terms can be neglected.  Similarly, for
small $h$, we find that $F$ is just the positive wavevector part of $h$.
At finite $\eta$, the equations of motion in terms of $h$ will
become singular due to the presence of overhangs, but
the equation of motion in terms of $F$ will not have any such problems.

At given $\eta$, the dynamics of the cluster can be obtained from the
Shraiman-Bensimon equation\cite{shraibens}:
\be
\label{sb}
\partial_t F(x,t)=i(\partial_x F(x,t))\int {\rm d}x' \,
|\partial_{x'}F(x')|^{-1-\eta} \frac{e^{ix'}+e^{ix}}{e^{ix'}-e^{ix}}
\ee

Let us introduce a regularization into the equation.  We would like
to add a term that cuts the problem off at a constant length in real
space, corresponding to the finite walker size in a DLA or DBM problem.
However, for small $F(j)$, the constant cutoff in real space can be
replaced by a constant cutoff in $x$ space.
A constant cutoff in $x$ space can be obtained by writing
\be
\label{regular}
\partial_t F(x,t)=i(\partial_x F(x,t))\int {\rm d}x' \,
|\partial_{x'}F(x'+i\eta \nu)|^{-1-\eta} \frac{e^{ix'}+e^{ix}}{e^{ix'}-e^{ix}}
\ee
for some $\nu$.  While equation (\ref{sb}) leads to finite time singularities,
equation (\ref{regular}) does not.

Let us write this equation in terms of $F(j)$ and
expand this equation in powers of $\eta$ and $a_k$.
The linear term is
\be
\label{lin}
\partial_t F(j)=\eta k F(k)-\eta \nu k^2 F(j)
\ee
Unlike the linear term,
the nonlinear terms in equation (\ref{sb}) are nonvanishing in the limit
$\eta \rightarrow 0$.  
The first nonlinear term is, at $\eta=0$,
\be
\label{nlin}
\partial_t F(j)=-\sum\limits_{k>0} \Bigl(k(j-k) F(k) F(j-k)-
2k(j+k) \overline F(k) F(j+k)\Bigr)
\ee
There is also a constant term in the expansion of equation (\ref{regular})
which simply describes an overall upward motion of the surface.  We will
drop this term by going to the comoving frame of the surface.

For small $\eta$, so long as the combination of equations (\ref{lin},\ref{nlin})
leads to well behaved solutions, we expect that $F(k)$ will be of order
$\eta$, simply by scaling.  This implies that all higher order non-linearities
will be unimportant in the limit.  Since $F(k)$ is taken to be of 
order $\eta$, the fluctuations about
the flat surface vanish in the limit of vanishing $\eta$, 
and so the growing cluster has fractal dimension 2
in this limit, as explained in the introduction.

So, if we rescale the field $F$ by $\eta$, and also rescale the time
coordinate, the equation of motion for the system will be
\be
\label{eqm}
\partial_t F(j)=k F(k)-\nu k^2 F(j)
-\sum\limits_{k>0} \Bigl(k(j-k) F(k) F(j-k)-
2k(j+k) \overline F(k) F(j+k)\Bigr)
\ee

While the full equation (\ref{regular}) leads to dynamics without
finite-time singularities, we have not shown this for equation (\ref{eqm})
obtained from equation (\ref{regular}).  However, all of the solutions
we write down are well-behaved for all time, so we will not worry about
this point too much.

Using equation (\ref{height}) and equation (\ref{eqm}) we can
obtain a
differential equation describing the growth of the height.  The result is
\be
\label{diffeq}
\partial_t h(x,t)=|\partial_x| h(x,t) + \nu \partial_x^2 h(x,t) +
(\partial_x h(x,t))^2
\ee
This is equivalent to equation (\ref{eqm}).
This equation is similar to a KPZ equation\cite{KPZ}, except for the presence of
the term $|\partial_x| h(x,t)$, which leads to an instability.
The operator $|\partial_x|$ is a non-local operator in real space, equal to
$|j|$ in Fourier space.
The presence of this term change the dynamics greatly from the KPZ problem.
In the KPZ problem, all states eventually relax to the state 
$h(x,t)={\rm const}$. in the long time, zero noise, limit.  The instability
implies that the state with constant $h(x)$ is not stable to small 
perturbations.  

Let us discuss the possibility of perturbing equation (\ref{diffeq}) with
noise.  We can easily modify the equation to
\be
\partial_t h(x,t)=|\partial_x| h(x,t) + \nu \partial_x^2 h(x,t) +
(\partial_x h(x,t))^2+{\rm noise}
\ee
where some random noise field has been added.  We expect that the noise will
be short range correlated on a length and time scale set by the short-distance
regularization.  However, the magnitude of the noise should also depend
on the short distance length scale.  The physical origin of the
noise is shot noise due to fluctuations in the number of walkers arriving
at a given point; the size of these fluctuations will depend on the
size of the walkers, as for a large number of walkers the shot noise will
get averaged out.  To get at the magnitude of the
noise, let us discuss some scaling properties of equation (\ref{diffeq}).

We assumed from the beginning boundary conditions such that $x$ was
periodic with period $2\pi$.  To discuss scaling, we will change these
boundary conditions; we will now measure $x$ in terms of a short-distance
length scale set by the regularization, so that $x$ is periodic
with period $2\pi/\nu$.  The natural differential equation
to describe the dynamics at short distances, of order the walker size, is
\be
\label{bare}
\partial_t h(x,t)=|\partial_x| h(x,t) + \partial_x^2 h(x,t) +
(\partial_x h(x,t))^2+N(x,t)
\ee
where the correlation length and time of the noise field $N(x,t)$ are now of 
order unity, and the mean-square fluctuations in $N(x,t)$ are also of
order unity.  The magnitude of the fluctuations will be set by noise-reduction
parameters\cite{noisered}, but will not depend on the macroscopic length
scale, $2\pi/\nu$.
Now, let us rescale length and time by a factor of $\nu$, while rescaling
$h(x,t)$ by a factor of $1/\nu$.  The result is
\be
\label{irlimit}
\partial_t h(x,t)=|\partial_x| h(x,t) + \nu\partial_x^2 h(x,t) +
(\partial_x h(x,t))^2+\tilde N(x,t)
\ee
where $\tilde N$ is short-range correlated with mean-square fluctuations
of order $\nu^2$.
Equation (\ref{irlimit}) is the same as equation (\ref{diffeq}), with an 
appropriate value of
noise added.  As $\nu$ gets smaller, meaning that the ratio of the IR 
cutoff to the UV cutoff gets larger, implying that we are approaching the
true long-distance behavior, we see that the terms
$\nu \partial_x^2 h(x,t)$ and $\tilde N(x,t)$ get smaller, suggesting that
these terms could be dropped in the IR limit.  However, the term
$\nu \partial_x^2 h(x,t)$ is dangerously irrelevant, so it would be too
quick to assume that this term (or the noise term) could be dropped; in
fact, we will find later that the noise becomes very important in
the IR limit.
\section{Pole Dynamics at Non-Vanishing $\nu$}
We will look for solutions to equation (\ref{diffeq}).  The natural way to
describe these solutions is in terms of the poles of $\partial_x F(x,t)$.
First we will find a solution with one pole and  then describe the dynamics of
a solution with several poles.  This dynamics is similar to
Calogero-Sutherland dynamics, but different enough that the dynamics of
the system of poles is not integrable.  At the end of this section, we also 
consider
the system with different boundary conditions, so that $F(x,t)$ is defined
for all $x$ on the real line, instead of just on a strip of width $2\pi$.
The pole dynamics takes a slightly simpler form in that case.

Similar pole decompositions have been found before, in the KPZ
equation\cite{kppole} and KdV equation\cite{kdpole}.  The pole decomposition
in continuum DLA is also well known\cite{shraibens,srev}.  The interest in the
particular pole decomposition here is that the equation of motion we
are considering is regularized (unlike that in continuum DLA) {\it and}
is unstable when considered near $h(x)=0$ (unlike KPZ).

First let us give a simple solution to the equation of motion.  From
the scaling arguments at the end of the last section, we can guess that
we want $F(k)$ to behave as $1/k$ for small $k$.  The solution we will
pick is
\be
\label{trial}
F(k)=-\frac{\nu}{k}e^{-i\epsilon k}
\ee
for $k>0$, with $\epsilon$ being negative imaginary.
One may verify that, if equation (\ref{trial}) is inserted into
equation (\ref{eqm}), the result is
\be
\partial_t \epsilon=i -i \nu -2i \nu \frac{e^{2\,{\rm Im}(\epsilon)}}
{1-e^{2\,{\rm Im}(\epsilon)}}
\ee
We find that if $\epsilon=-i{\rm log}(\frac{\sqrt{1+\nu}}{\sqrt{1-\nu}})$
then $\partial_t \epsilon=0$ and we have a stationary solution to the
equation of motion.
In the small $\nu$ limit, $\epsilon \approx -i\nu$.

From equation (\ref{trial}), we find that
\be
\partial_x F(x)=-i\nu\frac{e^{-i\epsilon+ix}}{1-e^{-i\epsilon+ix}}
\ee
In the small $\epsilon$ limit this reduces to
\be
-i \nu \frac{1}{i\epsilon-ix}
\ee
So, $\partial_x F(x)$ has a pole at $x=i\epsilon$.
Also we find that
\be
F(x)= \nu {\rm log}(1-e^{-i\epsilon+ix})
\ee
\be
\label{contrib}
h(x)=\frac{\nu}{2}
{\rm log} \Bigl(1+e^{2\,{\rm Im}(\epsilon)}-2 e^{{\rm Im}(\epsilon)}
{\rm Cos}(x-{\rm Re}(\epsilon))\Bigr)
\ee
Therefore, for $x$ near ${\rm Re}(\epsilon)$, the height $h(x)$ has
a very deep depression.  The function $h(x)$ is negative and diverging 
logarithmically (up to a cutoff set by ${\rm Im} \epsilon$) as $x$ approaches
${\rm Re}(\epsilon)$.  For $x$ near ${\rm Re}(\epsilon)+\pi$, the
function $h(x)$ has a very broad finger.  One should expect that the
fingers would be broad in this problem, as $\eta$ is taken to be infinitesimal
and we know physically that at smaller $\eta$ the fingers get wider.
A picture of $h(x)$ for this single pole solution is shown in figure 1.

Under the dynamics, the position of the logarithm in $h$ moves, but
its magnitude does not.  This is interesting, since it suggests that
$e^{h(x)/{\nu}}$, a quantity that arises in a Cole-Hopf transformation
of Burgers equation, has simple zeroes.

We can consider a more complicated solution combining $N$ different poles.
Consider
\be
\label{trial2}
F(k)=-\sum_{a=1}^{N} \frac{\nu}{k}e^{-i \epsilon_a k}
\ee
In this case, the equation of motion yields
\be
\label{eqmp}
\partial_t \epsilon_a=i-i\nu-2i \nu \Bigl(\frac{e^{2\,{\rm Im}(\epsilon_a)}}
{1-e^{2\, {\rm Im}(\epsilon_a)}}
+\sum_{b\neq a} (\frac{1}{1-e^{i\epsilon_b-i\epsilon_a}}+
\frac{1}{e^{i \epsilon_a-i \overline \epsilon_b}-1})\Bigr)
\ee

The dynamics of equation (\ref{eqmp}) causes poles to attract in
the real direction while repelling in the imaginary direction.  Consider
a problem with only two poles.  If the two poles
are nearby, with the same imaginary coordinate, but different real
coordinates, the poles approach
and collide in a finite time. 
However,
this singularity causes no problem: the poles hit, and then move away
from each other in the imaginary direction.  Until the poles collide, both
poles have the same imaginary coordinate and different real coordinates.
After collision, they have the same real coordinate and different
imaginary coordinates.  If the initial imaginary coordinates are
even infinitesimally different, the poles never collide.
The motion of this two pole problem is sketched
in figure 2.

The physical mechanism for the attraction of the poles in the real
direction is simple.  As discussed above, a given pole causes
a large depression in the surface.  The surface slopes down towards
this depression on both sides.  This 
gradient in $h$ then attract the other poles.

It is interesting to note that the attraction of poles in the real
direction provides a physical interpretation for the fact that each
pole contributes an amount of order $\nu$ to $h(x)$ (see
equation (\ref{contrib}).  Each pole
can also be thought of as having a ``self-attraction".  The term
$\partial_x^2 h(x)$ tends to smear out poles in the real direction;
only if a pole has weight at least $\nu$ can it attract itself
enough to avoid being spread out.  The term in the nonlinear
equation of motion which causes the interaction between poles is
exactly the same as the term that cancels $\partial_x^2 h(x)$ in the
single pole dynamics.

Without the term $-i+i\nu$ in equation (\ref{eqmp}), the equation could
be reduced to a complexification of
Calogero-Sutherland dynamics\cite{kppole,csuth}, and then
the pole dynamics would be integrable.  It seems to us that the
term $-i+i\nu$ destroys integrability in equation (\ref{eqmp}).

The dynamics of equation (\ref{eqmp}) assumes a simpler form if we
assume that the poles are all very near to each other with $\epsilon$
and $\nu$
small, or, equivalently, if
we change boundary conditions so that $x$ is now defined on the real line
and is not periodic.  Then we find
\be
\partial_t \epsilon_a=i-2\nu \Bigl(\frac{1}
{2i\,{\rm Im}(\epsilon_a)}
+\sum_{b\neq a} (\frac{1}{\epsilon_a-\epsilon_b}+
\frac{1}{\epsilon_a-\overline \epsilon_b})\Bigr)
\ee

\section{Stationary States of Poles}
In this section, we consider the qualitative features of the pole dynamics
found in the previous section.  First, we consider the case with $F(x,t)$
defined on the real line, and find the long time limit of the pole dynamics,
for any finite number of poles.  It is found that, for given number of
poles, the long time limit of $F(x,t)$ is independent of the initial positions
of the poles, up to a trivial translation in $x$.  Then, we consider the case 
with periodic boundary conditions, and find that, for given initial
number of poles, the long time limit of $F(x,t)$ tends to one of a discrete
number of different states, again up to translations in $x$.  
It is argued that all except for one of these states is unstable to small
perturbations in the initial configuration of poles.

As discussed in the previous section, the poles attract each other in
the real direction, and repel in the imaginary direction.
If we have a finite number of poles, with $F(x,t)$ defined on the real
line, it is clear that, in the long time limit, all poles will have the
same real coordinate (one can easily show this from
the equation of motion by taking a time derivative of $\sum_a
({\rm Re}(\epsilon_a))^2$).
Although the poles repel each other in the imaginary direction,
they are attracted to the real axis.  The result is
that the poles form some stable configuration with all poles having the
same real coordinate, and with different imaginary coordinates.
There is only one stationary configuration for a given number of poles.
This configuration is stable in that any small
perturbation of the pole positions will decay in time, except for
the trivial zero mode associated with translation.
If this configuration has $N$ poles, then $h(x)$ behaves as
$\nu N {\rm log}(x)$ for large $x$, so configurations with different
numbers of poles have different boundary conditions at
large $x$.

Consider stationary states of poles with periodic boundary conditions.  One
possibility is for all poles to have the same real coordinate, up
to multiples of $2\pi$.
There are, however, other possibilities.  One very simple possibility is
to take 2 poles, with ${\rm Re}(\epsilon_1)=0$ and
${\rm Re}(\epsilon_2)=\pi$.  In this case, the poles can find a stationary
configuration, but this configuration will not be stable: any small 
perturbation in the real part of one of the $\epsilon_a$ will cause
the two poles to move towards each other, until a new, stable, configuration
is found with both poles having the same real part of $\epsilon$.
This process is exactly the process of competition between two fingers,
with one finger eventually winning out, commonly seen in DLA and Hele-Shaw
problems.  We show some stages of this process in figures 3 and 4.
There are of course more complicated possibilities for stationary states,
but the only stable states are those with all poles having the same real 
coordinate.

Let us show for strip boundary conditions that there is a maximum
number of poles one can use to form these stable configurations.  
Consider what equation (\ref{eqmp}) implies for such a configuration.
Clearly, ${\rm Re}(\partial_t \epsilon_a)=0$ for all $a$.  We also need
${\rm Im}(\partial_t \epsilon_a)=0$ for all $a$.  Let us order the poles so
that if $a<b$ then ${\rm Im}(\epsilon_a)>{\rm Im}(\epsilon_b)$.  Consider
the equation of motion for $\epsilon_N$.  We find that
\be
\label{eqnN}
\partial_t \epsilon_N=i-i\nu-i 2\nu \Bigl(\frac{e^{2\,{\rm Im}(\epsilon_N)}}
{1-e^{2\,{\rm Im}(\epsilon_N)}}
+\sum_{b\neq a} (\frac{1}{1-e^{i\epsilon_b-i\epsilon_N}}+
\frac{1}{e^{i \epsilon_N-i \overline \epsilon_b}-1})\Bigr)
\ee
The first term on the right-hand side is positive imaginary, while all the
other terms are negative imaginary.  So long as 
${\rm Im}(\epsilon_N)<{\rm Im}(\epsilon_a)$ for all $a\neq N$, then
the imaginary part of the
right-hand side is monotonically increasing
as ${\rm Im}(\epsilon_N)$ becomes more and more negative.
In the limit $\epsilon_N\rightarrow -i\infty$, the
right-hand side becomes
\be
i(1-\nu -2 \nu (N-1))
\ee
If this quantity is negative imaginary, then it is not possible, for any
$\epsilon_N$, to have $\partial_t \epsilon_N=0$.  Then, $\partial_t
{\rm Im}(\epsilon_N)<0$ for all time, and we find that $\epsilon_N$ moves off to
$-i\infty$.  In the limit that $\epsilon_N$ moves off to infinity, this
pole no longer contributes to the sum in equation (\ref{trial2}) and can
be ignored.

Putting all these arguments together, we find that, for any $\nu$ in
the strip geometry, we can find states with $N$ poles, stable against
any small perturbation in the pole position.  For each $N$, there
is only one such state, and it has all poles with the same real
coordinate.  Such a stable solution with $N$ poles can only be constructed if
$-1+\nu +2 \nu (N-1)<0$.  Otherwise, some number of poles move off to
infinity until the solution reduces to one with $-1+\nu +2\nu(N-1)<0$.
\section{Stability Analysis}
We will look at the linear stability of the stationary states found
in the last section, for the case of periodic boundary conditions.
Here, we will 
consider arbitrary small perturbations of $h$, including perturbations
that cannot be written in terms of a small change in the pole
coordinates.
We are able to analytically calculate the eigenvalues of the linear
problem, and find that, for each value of viscosity, only one of the stationary
states is linearly stable. 
After obtaining the result, it is argued that
the qualitative features of the result could be obtained from the analysis
of the stationary states in the previous section: the linearly stable state is
the only state stable towards adding a pole.

Before doing any mathematical analysis of our problem, 
let us recall the stability analysis of the Saffman-Taylor system\cite{safft}.
Saffman and Taylor found a one-parameter family of stationary solutions
to the zero surface-tension problem.  Surface tension acts as a singular
perturbation to this problem.  Eventually it was shown\cite{bensimon} that
a small amount of surface tension stabilizes the ``1/2" finger against
small perturbations.  However, in the DLA problem, although the system is
regularized due to non-vanishing walker size, 
a fractal structure emerges instead of stable fingers.  In our
system, although the system is stable, in the IR limit an
exponentially small amount of noise will destabilize it, as
will be discussed more in section VII.  In that
case, it may turn out that we obtain a complicated structure in our
problem, for a small amount of noise.

The physical reason for stability is most easily understood in a WKB
approximation.  Consider a localized, short-wavelength, perturbation $f(x)$ to
a stationary state, $F_0(x)$.  Let the stationary state be such that all
poles have imaginary coordinate 0.  Then, the surface has a deep
depression near $x=0$, and a very broad finger located around $x=\pi$.
The perturbation will move in position,
due to the term in the equation of motion $\partial_x f(x) \partial_x F_0(x)$,
until the perturbation moves to $x=0$, where it disappears into the
deep depression on the surface.  A similar physical mechanism for stability
is known in the Hele-Shaw problem and other problems\cite{wkbk}.  The
closer the perturbation is to the tip of the finger initially, the
longer it will take it to drift along the side of the finger and
disappear.  Fortunately, the non-vanishing viscosity prevents us from
localizing a perturbation exactly at the tip as the perturbation
must have a width of order $\nu$ or greater if it is to be
unstable, so eventually all short-wavelength perturbations will be 
destroyed.

Now, let us proceed to a more careful analysis.
Let $F_0(x)$ be a stationary state, and let us consider a small perturbation
$f(x)$.  We will consider the evolution of the system in time, to
linear order in $f$.  One possibility to do this is, of course, to
take the equation of motion (\ref{eqm}), writing $F(x)=F_0(x)+f(x)$, and
directly derive the equation of motion to linear order in $f$, obtaining 
$\partial_t f(x,t)=L f$, where $L$ is some linear operator whose
eigenvalues describe the stability of the state $F_0$.  However, this procedure
would be very awkward, and we will use a more clever technique.

Let $F_0(x)$ be a state formed using $N$ poles, at positions $\epsilon_1,
\epsilon_2,...,\epsilon_N$.  We will write a state $F(j)$, near to $F_0(j)$,
as
\be
F(j)=F_0(j)+\frac{f(j)}{j}-i\sum \limits_{a=1}^{N} \nu f_a e^{-i\epsilon_a j}
\ee
where
$f(j)$ is a function of $j$ and the various $f_a$ are numbers
representing small deviations of the pole coordinates from their original
positions.  By doing this, we have introduced some redundancy in describing the
possible perturbations to $F_0$.

If $f(j)=0$, then we can derive a linear equation of motion for
the $f_a$, simply using the pole dynamics equation.  We must place
the $N$ poles at positions 
$\epsilon_1+f_1,\epsilon_2+f_2,...,\epsilon_N+f_N$ and
linearize equation (\ref{eqmp}) about $f_a=0$.  
We will obtain some equation of the form
\be
\label{linpole}
\partial_t f_a=L_{a,b} f_b + M_{a,b} \overline f_b
\ee
for some linear operators $L,M$, with an implied sum over $b$.
The result of this
is that there are $2N$ different modes (there are $N$ complex
coordinates $f_a$).  One of these modes is a zero mode; 
this is a mode with all $f_a$ equal to
the same real constant.  From the analysis of last section,
if all $\epsilon_i$ have the same imaginary coordinate, then the other
$2N-1$ modes are all stable (have negative eigenvalue).

Now, consider the case with non-zero $f(j)$.  For simplicity, first
consider a case with initial conditions such that
$f(1)$ is non-zero, but all other $f(j)$ are zero.
Then, the linearized equation of motion yields
\be
\label{linstab}
\partial_t (F(j)-F_0(j))=(1-\nu 1^2) \delta_{j,1} f(1) + 2 \nu
\sum\limits_{i=1}^{N} 
\Bigl((1-\delta_{j,1})
f(1) e^{-i\epsilon_i (j-1)}+\overline {f(1)} e^{-i\epsilon_i (j+1)}\Bigr)
\ee
The reason for the factor of $(1-\delta_{j,1})$ in the sum is that
in equation (\ref{trial}), each pole contributes only to $F(k)$ for
$k>0$.
It is convenient now to rewrite equation (\ref{linstab}) as
\be
\partial_t (F(j)-F_0(j))=(1-\nu 1^2-2\nu N) \delta_{j,1} f(1) + 2 \nu
\sum\limits_{i=1}^{N} 
\Bigl(f(1) e^{-i\epsilon_i (j-1)}+\overline{f(1)} e^{-i\epsilon_i (j+1)}\Bigr)
\ee
which, combined with equation (\ref{linpole}), is equivalent to the equations
\be
\partial_t f(1)=(1-\nu 1^2 -2\nu N) f(1)
\ee
\be
\partial_t f_a=-2 i
(f(1) e^{i\epsilon_a}+\overline {f(1)} e^{-i\epsilon_a})
+L_{a,b} f_b + M_{a,b} \overline f_b
\ee

If we extend this procedure to the case of general $f(j)$, we find
\be
\partial_t f(j)=(j-\nu j^2 -2 j \nu N) f(j) +2 i \nu \sum \limits_{k>j}\sum
\limits_{a=1}^{N} \Bigl(
e^{i(k-j)\overline \epsilon_a}-e^{-i(k-j)\epsilon_a}
\Bigr)
f(k) 
\ee
\be
\partial_t f_a=-2 i\sum\limits_{j} 
(f(j) e^{i j\epsilon_a}+\overline {f(j)} e^{-i j\epsilon_a})+
L_{a,b} f_b + M_{a,b} \overline f_b
\ee

The above two equations fully define the linear evolution of the system.  Note
that the matrix describing the linear evolution of the system is 
{\it triangular}: we find that
$\partial_t f(j)$ depends only on $f(k)$ for $k \geq j$.
This makes it possible to directly read off the eigenvalues of the
matrix.  There are $2N$ eigenvalues which are just the eigenvalues
from the evolution of equation (\ref{linpole}).  Then there are eigenvalues
which are $j(1-2 \nu N)-\nu j^2$ with $j=1,2,3,...$  Each of these
eigenvalues must in fact be counted twice, since there is one such eigenvalue
with $f(j)$ pure real, and one such with $f(j)$ pure imaginary.
So long as
we consider the stationary state found in the last section, with $N$ poles,
all with the same imaginary coordinate and with 
$N$ the largest integer less than $\frac{1+\nu}{2\nu}$, 
then all eigenvalues of the linear
evolution are negative, with one (or three) exceptions.  The first exception
is the zero mode corresponding to changing the imaginary coordinate of all
the poles by the same amount.  The other two possible exceptions occur if
$\frac{1+\nu}{2\nu}$ is an integer, in which case
$1-2\nu N-\nu 1^2=0$, and so we have two more zero modes.  However,
in any case, there are no positive eigenvalues.

The above analysis might be slightly confusing, as it seems we have introduced
more eigenvectors than we started with by adding the $f_a$ coordinates.
If we look at the linearized equation of motion in its original form, without
introducing the additional coordinates $f_a$, we notice that the diagonal term
in the linear equation of motion is $j-\nu j^2$.  For very large $j$, this
term must dominate all other terms, and so, if we order the eigenvalues
of the linearized equation of motion, and look at the $2j$-th eigenvalue,
this eigenvalue must be close to $j-\nu j^2$.  This gives us one
way to count the number of eigenvalues of the problem: since the dimension
of the space of $f(j)$ is infinite, we cannot simply count the number of
eigenvalues directly, but we can count the number of eigenvalues less than
a given number.
Now, notice that
\be
j(1-2 \nu N)-\nu j^2=k-\nu k^2 -N+N^2\nu
\ee
with 
\be
\label{shift}
k=j+N
\ee
The quantity $-N+N^2 \nu$ will be of order $N$, but for very large $k$
it will be small compared to $k-\nu k^2$.
So, while we have introduced $2N$ eigenvalues by adding the $f_a$ coordinates,
equation (\ref{shift}) makes it clear that we have made up for this by
losing $2N$ eigenvalues elsewhere: the full set of eigenvalues includes
the $2N$ eigenvalues from equation (\ref{linpole}) as well as eigenvalues
$j-\nu j^2 -N+N^2\nu$ with $j=N+1,N+2,N+3,...$.  Then, if we look at the
$2j$-th eigenvalue, it will be close to $j-\nu j^2$.
It is possible to make this argument more precise to show that we have
found all eigenvalues of the linear evolution.

The linearly stable configuration is also stable to adding a pole.
If a pole is added, one pole will move off to $-i\infty$, and the
solution will be unchanged.  This provides some insight into the
reason for the linear stability of the solution.  One way to create
an infinitesimal perturbation to $h$ using only poles is to add
an additional pole at a very large negative imaginary coordinate.  If
$N$ is large enough, the added pole will move to infinity;
otherwise, the added pole will approach the
real axis and $F$ will not be stable to this perturbation.
\section{Fluid Dynamics}
Consider the limit of small $\nu$.  From equation (\ref{trial2}), each
pole contributes an amount of order $\nu$ to $F(x)$.  However, from the
analysis of the previous two section, we expect that the only stable
state is a state with $N$ poles, where, for small $\nu$, we
find $N \approx \frac{1}{2\nu}$.  So, in the limit of small $\nu$, we
will find that $F(x)$ tends to a finite, non-vanishing function.
We expect that, in the limit of vanishing $\nu$, the problem can be described
in terms of a ``fluid" of poles.  Let us analytically find the state
of this fluid which corresponds to the small $\nu$ limit of the stable
state found in the last section.

For finite $\nu$ and $N$, taking all poles at the same real coordinate,
the equation of motion (\ref{eqmp}) yields $N$ different equations which
constrain the imaginary coordinates of the poles.  These equations are
non-linear, and not easy to solve.  However, for small $\nu$, the number
of poles becomes large, while the separation of poles, and the contribution
of each pole to the right-hand side of equation (\ref{eqmp}), becomes small.
In that case, we can rewrite (\ref{eqmp}) in terms of a density of poles.

First, let us write the equation for the pole density in general, for
any state, not just a stationary state.  Let the pole density be
a function $\rho(\epsilon)$, defined such that
\be
\label{trial3}
F(k)=\int {\rm d}^2\epsilon \, \rho(\epsilon) e^{-i\epsilon k}
\ee
We find that the equation of motion (\ref{eqmp}) becomes
\be
\partial_t \rho(\epsilon)=-{\rm Re}(\overline\partial_\epsilon 
\Bigl(\rho(\epsilon) J(\epsilon)\Bigr))
\ee
where
\be
J(\epsilon)=i-2 i\int {\rm d}^2 \epsilon' \, \rho(\epsilon')
\Bigl(\frac{1}{1-e^{i\epsilon'-i\epsilon}}+
\frac{1}{e^{i\epsilon-i\overline \epsilon'}-1}\Bigr)
\ee
This can be found either as a limit of equation (\ref{eqmp}), or
by inserting equation (\ref{trial3}) into the zero viscosity limit of
equation (\ref{eqm}).  Here $J(\epsilon)$ represents the velocity at
which the fluid moves, so the equation of motion for the fluid is
just a continuity equation.

In the zero viscosity limit, it is not obvious
that $h(x)$ will always be well-defined.  
For finite viscosity, poles cannot hit the axis, but
for vanishing viscosity, if the fluid of poles moves
so that it intersects the real axis in a sufficiently singular
manner, equation (\ref{trial3}) will become ill-defined.  

Let us look for the stable configuration of poles.  Let all poles
have real coordinate equal to 0, so that $\rho(\epsilon)=
\delta({\rm Re}(\epsilon)) f(-{\rm Im}(\epsilon))$.
A stationary configuration of the
fluid is one such that $J(\epsilon)$ vanishes for any $\epsilon$ such
that $\rho(\epsilon)$ is nonvanishing. 
For the stable configuration, which
is the one with the maximum number of poles,
$\rho(\epsilon)$ is non-vanishing for all $\epsilon$ on the negative
imaginary axis; that is, there is
a density of poles extending off to infinity.

So, to find the stable configuration, we must find a function $f$ such that
$J$ vanishes on the negative imaginary axis.  
That is, we must solve the integral equation
\be
\label{integq}
-1+2\int\limits_0^{\infty} {\rm d} y \, f(y)
\Bigl(\frac{1}{1-e^{y-x}}+
\frac{1}{e^{x+y}-1}\Bigr)=0
\ee
for all real $x>0$.
Equation (\ref{integq}) can be rewritten as
\be
2\int\limits_0^{\infty} {\rm d} y \, f(y)
\frac{1}{{\rm Cosh}(x)-{\rm Cosh}(y)}=\frac{1}{{\rm  Sinh}
(x)}
\ee

Finally, introducing a new coordinate, $z={\rm Cosh}(x)$, and
$z'={\rm Cosh}(y)$, we find
\be
2\int\limits_1^{\infty} \frac{{\rm d} z'}{\sqrt{(z')^2-1}}
\, f(z')
\frac{1}{z-z'}=\frac{1}{\sqrt{z^2-1}}
\ee
for all $z>1$.
This is an integral equation which can be solved using known 
techniques for singular integral equations\cite{mushkelishvili}.

We find that
\be
f(x)=\frac{1}{2\pi^2} {\rm log}(\frac
{1+{\rm Cosh}(x)}{1-{\rm Cosh}(x)})
\ee
This function has vanishes exponentially for large $x$.  This
implies that the density of poles vanishes exponentially
for large, negative imaginary coordinates.

For this given $f$, one can examine the function $h(x)$ that results.
As $x\rightarrow 0$ one finds that $\partial_x h(x)$ diverges
logarithmically.  However, the height $h(x)$ does not diverge as 
$x\rightarrow 0$.  It is interesting that the fingering solution
in the zero viscosity limit is so well behaved for $\eta\rightarrow 0^{+}$; 
for $\eta=1$ it is known
that the stable finger only occupies half the width of the channel, going
off to infinity at two distinct points.  It is possible that, for
any non-zero $\eta$, the solution we have found will narrow so
that it does not occupy the whole channel.
\section{Noise in the IR Limit and Perturbation Theory}
From the naive scaling analysis of equation (\ref{bare}), we concluded in
section II that the noise vanishes in the IR limit.  However, as we noted, the
viscosity also vanishes in this limit, and since the viscosity
is a dangerously irrelevant operator, this can change everything.
In this section we will argue that, in fact, in the IR limit, the
system becomes unstable to an exponentially small amount of noise.
Then, we will briefly discuss the perturbation theory needed to deal
with finite $\eta$.  However, the lack of stability of the $\eta\rightarrow
0^{+}$ system
in the IR limit implies that we should not apply the perturbation
theory in $\eta$ until we have first found some way to deal with noise
at infinitesimal $\eta$.

Consider the
equation of motion (\ref{bare}), so that we are using the short-distance
length scale.  On this length scale,
the stationary configuration of $h(x)$, found above, appears to be very
smooth.  So, for the interaction of the stationary configuration with
the noise, we can use a WKB approximation, as discussed above.
In the WKB approximation, the perturbations to the surface move along the
surface until they hit the depression caused by the poles, where the
perturbations are destroyed.  

In the rescaled lengths, it takes a
time of order $v^{-1}$ to hit the depression.  So, any small perturbation
will have a long time to grow before it is
destroyed.  In rescaled lengths and times, the most unstable wavevector, 
when considering fluctuations
about a flat surface, is $j=1/2$, which grows at a rate
equal to $1/4$.  
So, a short distance perturbation to the surface
will grow at a rate of order $1/4$, for a time of order $1/\nu$, before
being destroyed.  So, the perturbation will be amplified by a factor 
such as $e^{\frac{1}{4\nu}}$.  So, an exponentially small amount of
noise will lead to non-linear instabilities of the surface.

As discussed
in section V, it is impossible to localize a perturbation exactly
at the tip of the finger.
However, the greatest growth will occur for perturbations located at the
tip of the finger, a distance of
$\pi$ away from the depression, as these perturbations will last the
longest before being swept into the depression. 
It is expected that perturbations near the tip lead to
a tip splitting instability.  

The exponential
non-linear instability can also be seen from the linear stability analysis
above; many of the eigenvectors are almost degenerate.  In forming the
upper triangular matrix used in the linear stability analysis, many of the
elements above the diagonal are exponentially large.  

Let us briefly discuss the theory at finite $\eta$.  One must extend equation
(\ref{diffeq}) to include higher terms in $\eta$ and higher non-linearities
in $F$.  This leads to a perturbative correction of equation (\ref{eqm})
in powers of $\eta$.
The perturbation theory should include three effects.  First, the stationary
configurations found in the last section should change even in
the absence of noise.
Secondly, when we drive the theory by a small amount of noise, we have
argued above that the configuration becomes unstable in the IR limit.
For non-vanishing $\eta$ the fluctuations in $h(x)$ will interact
differently than at vanishing $\eta$.  However, until we can fully
understand the limit of vanishing $\eta$, there is no reason to 
consider the perturbation theory in detail.  Thirdly, we argued
that in the limit of vanishing $\eta$ is was acceptable to replace the
cutoff of constant length in real space by a cutoff of constant
length in $x$ space, where $x$ is used to parametrize the surface.
This is not valid for finite $\eta$, and must be corrected.
\section{Conclusion}
In conclusion, we have considered the problem of the dielectric
breakdown model in the limit of vanishing $\eta$.  A dynamics in terms
of poles was found, and was used to find the stable configuration of
the system for different values of $\nu$.  In the limit of vanishing
$\nu$, this provides us with a fingering solution similar to the
fingering solution of the Saffman-Taylor model.  As expected,
the finger is much broader than the Saffman-Taylor finger.

It was argued that in the IR limit an exponentially small amount of noise
destabilizes the system.  This will be discussed will be discussed
more in a future work, where we will also elaborate more on the branching
properties of the surface.

Once the $\eta\rightarrow 0^{+}$ problem is solved in the presence of noise,
an even more interesting and difficult problem is to 
obtain perturbative results for the DBM in powers of $\eta$.
Since the problem for nonvanishing $\eta$ gives rise to fractal structures,
with nontrivial scaling, perturbative results for the DBM should
involve further IR divergences,  which would have to be handled by an RG 
procedure and
would give rise to the non-trivial dimensions.  
We hope that the work in this paper represents a first step towards solving
that full problem.

{\it Note Added:}
After this paper was first posted to cond-mat, Vincent Hakim pointed out
to me some closely related results
in the literature on the problem of flame fronts.
Equation (\ref{diffeq}) is well known in that problem, where it is referred to 
as Sivashinksy's equation\cite{siva}.  The pole dynamics was previously
identified also\cite{thual}.  It appears, however, that the
analytic calculation of the linear 
stability that appears in this paper is new, as
well as the connection between the exponentially large
off-diagonal elements of the stability matrix and the WKB stability analysis;
some numerical results on the stability of
the system are known\cite{olami}, and agree with the exact analytic calculation
here.  Further, the connection between the
DBM and Sivashinsky's equation also appears to be new.

\begin{figure}[!t]
\begin{center}
\leavevmode
\epsfig{figure=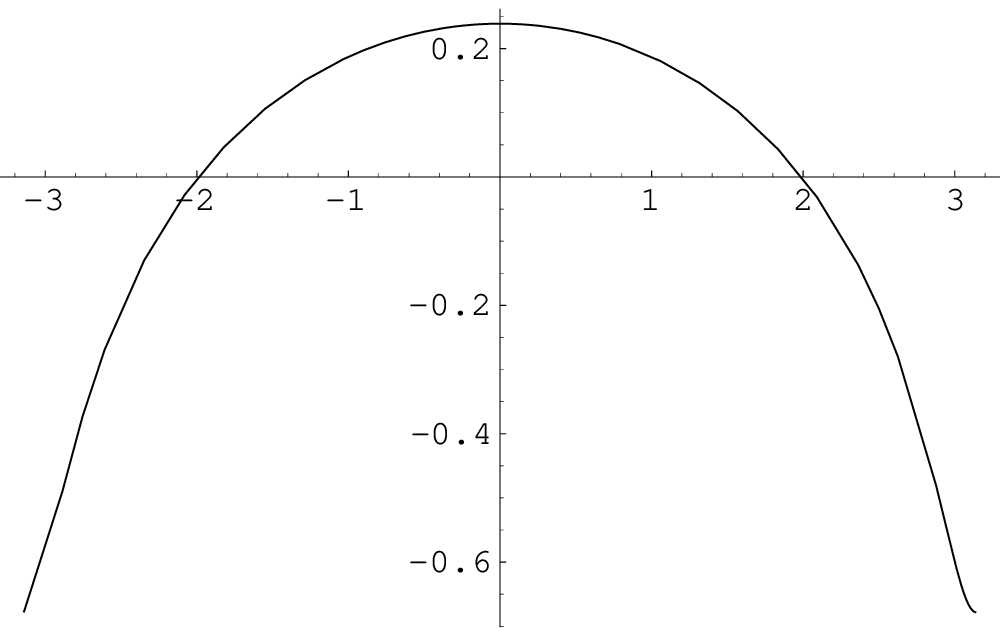,height=8cm,angle=0}
\end{center}
\caption{A plot of $h(x)$ for the trial solution at $\nu=.2$}
\label{fig1}
\end{figure}
\begin{figure}[!t]
\begin{center}
\leavevmode
\epsfig{figure=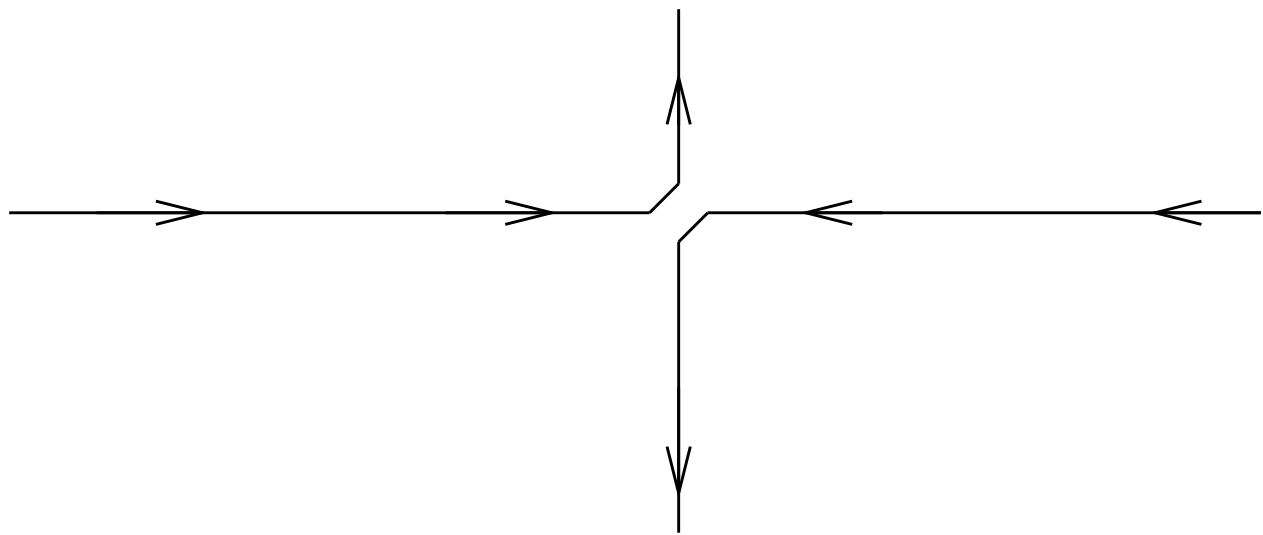,height=8cm,angle=0}
\end{center}
\caption{Sketch of paths of two poles.  Horizontal line represents
real axis; poles lie in lower half-plane.  Figure is approximate;  actual
paths will be rounded out.}
\label{fig2}
\end{figure}
\begin{figure}[!t]
\begin{center}
\leavevmode
\epsfig{figure=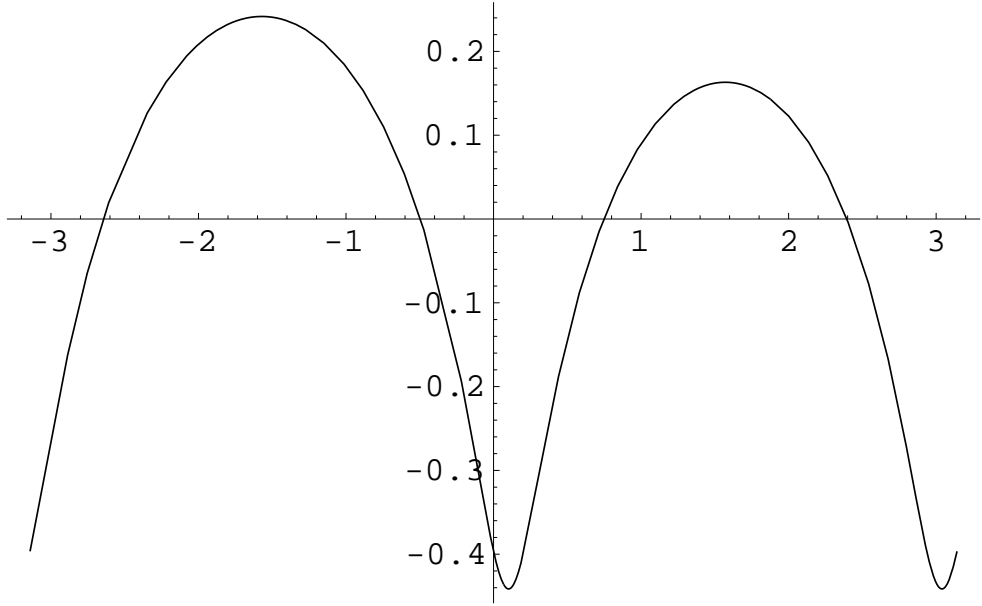,height=8cm,angle=0}
\end{center}
\caption{Competition between two fingers, short times.  Poles
have real coordinates $.1$ and $\pi-.1$.}
\label{fig3}
\end{figure}
\begin{figure}[!t]
\begin{center}
\leavevmode
\epsfig{figure=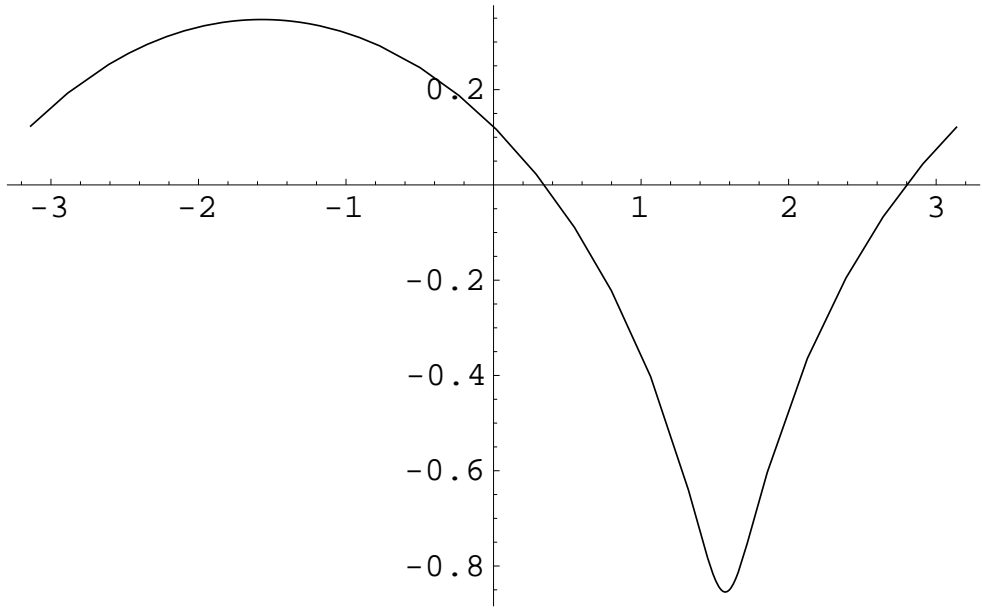,height=8cm,angle=0}
\end{center}
\caption{Competition between two fingers, long time limit where one
finger is completely suppressed.}
\label{fig4}
\end{figure}
\end{document}